\documentclass[letterpaper,10 pt,conference]{ieeeconf}
\IEEEoverridecommandlockouts 
\usepackage{amsmath,amssymb,amsfonts}
\usepackage{graphicx}
\usepackage{subfig}
\usepackage{textcomp}
\usepackage[T1]{fontenc}
\usepackage{color}
\usepackage{xcolor}
\usepackage{multirow}
\usepackage{multicol,lipsum}
\usepackage{epsfig}
\usepackage{algorithm}
\usepackage{algorithmic}
\usepackage{multirow}
\usepackage{hyperref}
\usepackage{mathtools}
\usepackage{bm}
\usepackage{listings}

\newtheorem{theorem}{Theorem}

\newtheorem{proposition}{Proposition}
\newtheorem{lemma}{Lemma}

\newtheorem{assumption}{Assumption}
\usepackage{array}
\usepackage{float}
\usepackage[short]{optidef}
\usepackage{mdframed}
\mdfdefinestyle{mystyle}{
    backgroundcolor=white!10}

\newcommand{\oomit}[1]{}

\newcolumntype{M}[1]{>{\centering\arraybackslash}m{#1}}
\newcolumntype{N}{@{}m{0pt}@{}}
\begin{document}
\title{Necessary and Sufficient Lyapunov-Like Conditions for Reach-Avoid Verification of Controlled Deterministc Discrete-Time Systems\thanks{Although the technical development may seem relatively straightforward from my perspective, I believe it is important to present the results and their derivation clearly, given the significance of the problem they address.}}

\author{Bai Xue \\\small Institute of Software, Chinese Academy of Sciences, Beijing, China 100190 \\
 xuebai@ios.ac.cn} 
\maketitle
\thispagestyle{empty}
\pagestyle{empty}

\begin{abstract}
Reach-avoid verification of controlled deterministic
discrete-time systems asks whether every initial state in a
given set admits an admissible control sequence that steers
the system to a target set while remaining within a prescribed
safe set until the target is reached. In this paper, we
establish necessary-and-sufficient Lyapunov-like
characterizations by combining a probabilistic reformulation
of controlled reach-avoid feasibility with converse results
for stochastic discrete-time systems. Specifically, we
derive characterizations in terms of lower-semicontinuous
certificate functions and continuous certificate functions under suitable regularity
assumptions. These results
provide a complete certificate-based characterization of
reach-avoid feasibility for controlled deterministic
discrete-time systems.
\end{abstract}

\section{Introduction}
Reach-avoid verification of controlled dynamical systems is a fundamental problem in formal methods and control theory, with particular importance for the rigorous assurance of learning-enabled autonomous systems. Given an initial set, a safe set, and a target set, the objective is to determine whether, from every initial state in the initial set, there exists a sequence of admissible control inputs that steers the system into the target set in finite time while keeping the state within the safe set until the target is reached. This problem arises naturally in safety-critical applications. For example, an autonomous vehicle must be able to reach its destination from a range of possible initial states while avoiding obstacles, and a robot must accomplish its assigned task without violating safety constraints \cite{Tomlin1999Computing,margellos2011hamilton}.

Inspired by classical Lyapunov methods for stability analysis
\cite{khalil2002nonlinear}, Lyapunov-like methods provide a
powerful framework for formal verification of dynamical
systems by reducing verification problems to the search
for certificate functions satisfying suitable inequalities
\cite{prajna2007framework,prajna2007convex}. These
inequalities can often be formulated as optimization
problems and, for polynomial systems, addressed using
sum-of-squares (SOS) programming
\cite{papachristodoulou2002construction}. While satisfying
such conditions provides sound verification guarantees,
failure to find a certificate does not necessarily imply
that the reach-avoid specification is violated, as it may
instead result from conservatism in the conditions. This
limitation motivates the study of necessary-and-sufficient
certificate characterizations, which establish an exact
correspondence between the system property and the
existence of a certificate under the stated assumptions
\cite{lin1996smooth,ratschan2018converse,liu2021converse,
mestres2025converse,majumdar2025sound,quartz2026converse,li2026converse}. Such characterizations are important for understanding
the completeness of Lyapunov-like verification methods
and providing a theoretical foundation for certificate-based
analysis. For controlled deterministic discrete-time systems,
however, an important open question remains: whether
reach-avoid feasibility admits a necessary-and-sufficient
characterization in terms of Lyapunov-like certificates. A promising approach to this question is to establish a
connection between controlled deterministic reach-avoid
verification and stochastic reach-avoid verification.

In our previous work \cite{wu2026controlled}, we developed
a probabilistic framework for computing controlled
reach-avoid sets (CRASs) of deterministic discrete-time
systems. By modeling control inputs as independent and
identically distributed random variables drawn from a
full-support probability distribution, we established
that deterministic reach-avoid feasibility is equivalent
to the corresponding stochastic reach-avoid event having
strictly positive probability. This reformulation enables
the use of Lyapunov-like conditions for stochastic systems,
leading to a convex optimization approach to CRAS
computation. It thus avoids the conventional joint
synthesis of control policies and certificate functions,
which generally results in a nonconvex optimization
problem. This equivalence provides a bridge between
controlled deterministic systems and stochastic systems,
opening the possibility of leveraging Lyapunov-like
characterizations for stochastic systems to address
controlled reach-avoid verification. This connection is particularly timely given recent advances in converse Lyapunov-like characterizations for stochastic discrete-time systems. Necessary-and-sufficient conditions have been established for infinite-horizon
reach-avoid verification \cite{xue2026sufficient} and
subsequently extended to uniform reach-avoid specifications
over compact initial sets \cite{xue2026converse}.
Furthermore, under suitable regularity conditions on the
transition kernels, these characterizations can be
established using continuous certificate functions
\cite{xue2026sufficient1}. Together, these developments
provide the key ingredients for extending converse
Lyapunov-like characterizations to controlled deterministic
discrete-time systems.

Building on these results, this paper establishes necessary-and-sufficient Lyapunov-like conditions for set-based reach-avoid verification of controlled deterministic discrete-time systems. By combining the probabilistic equivalence between controlled reach-avoid feasibility and positive-probability stochastic reach-avoid with converse characterizations for stochastic systems, we derive conditions based on lower-semicontinuous certificate functions. We further establish a continuous certificate characterization under a uniform absolute continuity condition on the induced transition kernels. These results extend converse Lyapunov-like theory from stochastic to controlled deterministic discrete-time systems.

The main contribution of this paper is therefore a complete Lyapunov-like characterization of reach-avoid verification for controlled deterministic discrete-time systems. Rather than introducing an entirely new certificate construction, we show how existing results on stochastic reach-avoid verification can be combined to establish this characterization.

Throughout this paper, we use the following basic notions: $\mathbb{R}$ denotes the set of real values; $\mathbb{N}$ denotes the set of nonnegative integers; for sets $\Delta_1$ and $\Delta_2$, $\overline{\Delta_1}$ denotes the closure of the set $\Delta_1$, $\Delta_1^{\circ}$ denotes the interior of the set $\Delta_1$, and $\Delta_1\setminus \Delta_2$ denotes the difference of sets $\Delta_1$ and $\Delta_2$, which is the set of all elements in $\Delta_1$ that are not in $\Delta_2$.  
\section{Preliminaries and Problem Formulation}
\label{sec:preliminaries}

This section introduces the controlled dterministic discrete-time system
model and formulates the reach-avoid verification problem
considered in this paper.

\subsection{Controlled Deterministic Discrete-Time Systems}

We consider a controlled deterministic discrete-time system of the form
\begin{equation}
    x(k+1)=f(x(k),u(k)), \qquad k\in\mathbb N,
    \label{eq:controlled-system}
\end{equation}
where $x(k)\in\mathbb R^n$ is the system state,
$u(k)\in\mathcal U\subseteq\mathbb R^m$ is the control input,
and $f:\mathbb R^n\times\mathcal U\to\mathbb R^n$ is the
state-transition function.

To describe the system evolution, let
$\pi=\{u(k)\}_{k\in\mathbb N}\in\mathcal U^{\mathbb N}$
denote an admissible control sequence. For a given initial
state $x_0\in\mathbb R^n$ and control sequence $\pi$, the
corresponding trajectory
$\phi_{x_0}^{\pi}:\mathbb N\to\mathbb R^n$ is defined
recursively by
\begin{equation}
    \begin{cases}
        \phi_{x_0}^{\pi}(0)=x_0,\\
        \phi_{x_0}^{\pi}(k+1)
        =f\bigl(\phi_{x_0}^{\pi}(k),u(k)\bigr),
        \qquad k\in\mathbb N.
    \end{cases}
    \label{eq:trajectory}
\end{equation}

\subsection{Reach-Avoid Specification}

Throughout the paper, we impose the following assumption
on the system dynamics and the sets involved in the
reach-avoid specification.

\begin{assumption}
\label{assump1}
The control set $\mathcal U$ is compact with $\mathcal{U}=\overline{\mathcal{U}^{\circ}}$ and has positive \(m\)-dimensional Lebesgue measure, and
the state-transition function $f$ is continuous in both
arguments. The safe set $\mathcal X\subseteq\mathbb R^n$
is bounded and open, the target set
$\mathcal T\subseteq\mathcal X$ is open, and the initial
set $\mathcal X_0\subseteq\mathcal X$ is compact.
\end{assumption}

Given an initial state $x_0\in\mathcal X$, the reach-avoid
objective is to determine whether there exists an admissible
control sequence that steers the system into the target set
$\mathcal T$ in finite time while keeping the state within
the safe set $\mathcal X$ until the target is reached.
Accordingly, we define the controlled reach-avoid set
(CRAS) as
\begin{equation*}
    \mathcal R
    :=
    \left\{
    x_0\in\mathcal X
    \;\middle|\;
    \begin{aligned}
        &\exists\,\pi\in\mathcal U^{\mathbb N},
        \ \exists\,k_0\in\mathbb N:
        \quad \phi_{x_0}^{\pi}(k_0)\in\mathcal T,\\
        &\phi_{x_0}^{\pi}(k)\in\mathcal X,
        \quad \forall k=0,\ldots,k_0
    \end{aligned}
    \right\}.
    \label{eq:cras}
\end{equation*}
Since $\mathbb N$ includes zero, an initial state already
in $\mathcal T$ satisfies the reach-avoid objective at
time zero.

The reach-avoid verification problem studied in this paper
is to determine whether every initial state in the prescribed
compact set $\mathcal X_0$ belongs to the CRAS, namely,
\begin{equation}
    \mathcal X_0\subseteq\mathcal R.
    \label{eq:reach-avoid-problem}
\end{equation}
Equivalently, this requirement can be expressed explicitly
in terms of the quantifiers over initial states, control
sequences, and target hitting times:
\begin{equation}
\begin{aligned}
    \forall x_0\in\mathcal X_0,\quad
    &\exists\,\pi\in\mathcal U^{\mathbb N},
    \quad \exists\,k\in\mathbb N:\\
    &\begin{cases}
    \phi_{x_0}^{\pi}(k)\in\mathcal T,\\
    \phi_{x_0}^{\pi}(i)\in\mathcal X,
    \quad \forall i=0,\ldots,k.
    \end{cases}
\end{aligned}
\label{eq:uniform-reach-avoid-problem}
\end{equation}

For subsequent certificate constructions, it is useful to
introduce a compact set containing the safe set and all
possible one-step successors of states in its closure.
Since $\mathcal X$ is bounded, $\overline{\mathcal X}$ is
compact. By the compactness of $\mathcal U$ and the
continuity of $f$, the image
$f(\overline{\mathcal X},\mathcal U)$ is also compact.
We therefore define the compact verification domain
\begin{equation}
    \mathcal X_b
    :=
    \overline{\mathcal X}
    \cup f(\overline{\mathcal X},\mathcal U),
    \label{eq:verification-domain}
\end{equation}
where
\[
    f(\overline{\mathcal X},\mathcal U)
    :=
    \{f(x,u) \mid x\in\overline{\mathcal X},
    \ u\in\mathcal U\}.
\]
By construction, $\mathcal X_b$ contains
$\overline{\mathcal X}$ and every one-step successor of
a state in $\overline{\mathcal X}$.

\section{Sufficient and Necessary Lyapunov-Like Characterization}
\label{sec:method}

Our approach proceeds in two stages. First, we reformulate the
controlled reach-avoid problem as a probabilistic one by
treating the control inputs as independent random variables
drawn from a full-support distribution. This reformulation
preserves reach-avoid feasibility and, together with the
compactness of the initial set, yields a lower-semicontinuous
Lyapunov-like characterization. We then impose an additional
regularity condition on the induced transition kernel to
establish the existence of continuous certificates.

\subsection{Probabilistic Reformulation}
\label{subsec:probabilistic-reformulation}

We associate a stochastic system with the controlled system
\eqref{eq:controlled-system} by treating the control inputs as
independent and identically distributed random variables:
\begin{equation}
    x(k+1)=f(x(k),\theta(k)),
    \qquad k\in\mathbb N,
    \label{eq:randomized-system}
\end{equation}
where $\{\theta(k)\}_{k\in\mathbb N}$ is an i.i.d. sequence
taking values in $\mathcal U$ according to a Borel
probability measure $\nu$. We impose the following
full-support assumption.

\begin{assumption}
\label{assum2}
The probability measure $\nu$ has full support on
$\mathcal U$, i.e.,
\begin{equation*}
    \nu(O)>0~\text{for every nonempty relatively open set }
    O\subseteq\mathcal U. 
    \label{eq:full-support}
\end{equation*}
\end{assumption}

A natural choice satisfying Assumption \ref{assum2} is the uniform
probability distribution on \(\mathcal U\).

Let $\boldsymbol\theta=(\theta(0),\theta(1),\ldots)$
denote a realization of the random input sequence, and let
$\mathbb P_\nu=\nu^{\otimes\mathbb N}$ be its product
probability measure on $\mathcal U^{\mathbb N}$. For an
initial state $x_0\in\mathcal X$, denote the corresponding
trajectory by $\phi_{x_0}^{\boldsymbol\theta}$. Throughout the paper, for a fixed state \(x\in\mathcal X\),
we use \(\mathbb E_\nu\) to denote expectation with respect
to the distribution \(\nu\) of a one-step random input
\(\theta\). Specifically, for a measurable function
\(g:\mathcal X_b\to\mathbb R\) for which the expectation
is finite, we define
\begin{equation}
    \mathbb E_\nu[g(f(x,\theta))]
    :=
    \int_{\mathcal U}g(f(x,u))\,\nu(du).
    \label{eq:one-step-expectation}
\end{equation}
Thus, $\mathbb E_\nu[g(f(x,\theta))]$ represents the
expected value of $g$ at the next state, with the current
state $x$ held fixed.

The reach-avoid hitting time is defined as
\begin{equation*}
\begin{aligned}
    \tau_{\mathrm{RA}}(x_0,\boldsymbol\theta)
    :=\inf\left\{k\in\mathbb N \middle |
    \begin{split}
    &\phi_{x_0}^{\boldsymbol\theta}(k)\in\mathcal T,\\
    &\phi_{x_0}^{\boldsymbol\theta}(j)\in\mathcal X,
    \quad j=0,\ldots,k
    \end{split}
    \right\},
\end{aligned}
\label{eq:stochastic-hitting-time}
\end{equation*}
where the infimum of the empty set is defined as
$+\infty$. The associated infinite-horizon reach-avoid
probability is
\begin{equation}
    P_{\mathrm{RA}}(x_0)
    :=
    \mathbb P_\nu\left(
        \tau_{\mathrm{RA}}(x_0,\boldsymbol\theta)<\infty
    \right).
    \label{eq:reach-avoid-probability}
\end{equation}

The following result, established in
\cite{wu2026controlled}, connects the deterministic and
probabilistic formulations.

\begin{proposition}
\label{prop:det-stoch-equivalence}
Under Assumptions 1 and 2, for every $x_0\in\mathcal X$,
\begin{equation}
    x_0\in\mathcal R
    \quad\Longleftrightarrow\quad
    P_{\mathrm{RA}}(x_0)>0.
    \label{eq:det-stoch-equivalence}
\end{equation}
\end{proposition}

In particular, the existence of an admissible control
sequence that achieves the reach-avoid objective from a
given initial state is equivalent to a strictly positive
probability of achieving the same objective under the
randomized system.  In \cite{wu2026controlled}, Lipschitz continuity of the system dynamics is used to establish the above equivalence.
However, this assumption can be weakened to continuity.
Indeed, if a deterministic control sequence achieves the
reach-avoid specification, the corresponding trajectory
satisfies the specification over a finite time horizon.
By continuity of the system dynamics and openness of the
safe and target sets, all control sequences in a sufficiently
small neighborhood of the corresponding finite control
sequence also satisfy the specification. The full-support
property of \(\nu\), together with the independence of the
randomized control inputs, then guarantees that this
neighborhood has positive probability. Conversely, positive
reach-avoid probability immediately implies the existence
of a deterministic control sequence satisfying the
specification. Thus, continuity of the system dynamics
suffices to establish the equivalence.

\subsection{Lower-Semicontinuous Lyapunov-like Characterization}
\label{subsec:lsc-characterization}

We first recall two results established in
\cite{xue2026converse}, which provide the basis for
deriving a lower-semicontinuous certificate
characterization. The first establishes the lower
semicontinuity of the reach-avoid probability.

\begin{lemma}[\cite{xue2026converse}]
\label{lem:lsc-probability}
Under Assumption~\ref{assump1}, the function
$P_{\mathrm{RA}}:\mathbb{R}^n\to[0,1]$ is lower
semicontinuous.
\end{lemma}

Together with the compactness of the initial set,
this regularity property allows pointwise positivity
of the reach-avoid probability to be strengthened
to a uniform positive lower bound.

\begin{proposition}[\cite{xue2026converse}]
\label{prop:uniform-margin}
Suppose that
\begin{equation}
    P_{\mathrm{RA}}(x)>0,
    \qquad \forall x\in\mathcal X_0.
\end{equation}
Then there exists $\varepsilon\in(0,1)$ such that
\begin{equation}
    P_{\mathrm{RA}}(x)>\varepsilon,
    \qquad \forall x\in\mathcal X_0.
    \label{eq:uniform-positive-margin}
\end{equation}
\end{proposition}

By Proposition~\ref{prop:det-stoch-equivalence},
controlled reach-avoid feasibility implies pointwise
positivity of the reach-avoid probability over
$\mathcal X_0$. Proposition~\ref{prop:uniform-margin}
then yields
\begin{equation*}
    \mathcal X_0\subseteq\mathcal R
    \quad\Longrightarrow\quad
    \exists\,\varepsilon\in(0,1):
    \quad
    P_{\mathrm{RA}}(x)>\varepsilon,
    \quad \forall x\in\mathcal X_0.
\end{equation*}
This uniform probability bound enables us to invoke the
necessary-and-sufficient lower-semicontinuous
Lyapunov-like characterization for stochastic
discrete-time systems established in
\cite{xue2026converse}. Applying this characterization
to the stochastic system induced by randomizing the
control inputs yields the following
lower-semicontinuous certificate characterization
for the controlled deterministic system
\eqref{eq:controlled-system}.

\begin{theorem}[\cite{xue2026converse}]
\label{thm:uniform-lsc}
Under Assumptions \ref{assump1} and \ref{assum2},
$\mathcal X_0\subseteq\mathcal R$ holds if and only if
there exist $\varepsilon\in(0,1)$, $\gamma\in(0,1)$\footnote{As shown in \cite{xue2026converse}, the parameter \(\gamma\) can be fixed a priori arbitrarily close to one without compromising the theoretical characterization. Thus, pre-specifying \(\gamma\) in practical computations is theoretically justified.},
and a lower-semicontinuous function
$v:\mathcal X_b\to\mathbb R$, bounded on $\mathcal X$,
such that
\begin{equation}
\label{eq:lsc-barrier-conditions}
\begin{cases}
    v(x)\geq\varepsilon,
    & \forall x\in\mathcal X_0,\\[2pt]
    v(x)\leq
    \gamma\mathbb E_\nu[v(f(x,\theta))],
    & \forall x\in\mathcal X\setminus\mathcal T,\\[2pt]
    v(x)\leq1,
    & \forall x\in\mathcal T,\\[2pt]
    v(x)\leq0,
    & \forall x\in\mathcal X_b\setminus\mathcal X.
\end{cases}
\end{equation}
\end{theorem}
\begin{proof}
By Propositions~\ref{prop:det-stoch-equivalence}
and~\ref{prop:uniform-margin}, the condition
$\mathcal X_0\subseteq\mathcal R$ is equivalent to the
existence of $\varepsilon\in(0,1)$ such that
\[
    P_{\mathrm{RA}}(x)>\varepsilon,
    \qquad \forall x\in\mathcal X_0.
\]
The necessary-and-sufficient lower-semicontinuous
characterization established in
\cite{xue2026converse} then yields the existence of
$\gamma\in(0,1)$ and a lower-semicontinuous certificate
satisfying \eqref{eq:lsc-barrier-conditions}, and
conversely.

As also noted in \cite{xue2026converse}, the certificate
can be restricted to the domain \(\mathcal X_b\) rather
than defined on the entire state space \(\mathbb R^n\).
For completeness, we briefly justify this restriction.  By definition
of $\mathcal X_b$, every one-step successor of a state in
$\mathcal X$ belongs to $\mathcal X_b$. Consequently,
the values of the certificate outside $\mathcal X_b$
do not affect any of the certificate inequalities.
Restricting a certificate on $\mathbb R^n$ to
$\mathcal X_b$ therefore preserves all the required
properties.

Conversely, suppose that
$v:\mathcal X_b\to\mathbb R$ satisfies
\eqref{eq:lsc-barrier-conditions}. Extend it to
$\mathbb R^n$ by setting
\[
    \bar v(x)=
    \begin{cases}
        v(x), & x\in\mathcal X_b,\\
        0, & x\notin\mathcal X_b.
    \end{cases}
\]
Since $v\leq0$ on $\mathcal X_b\setminus\mathcal X$,
the extension satisfies
$\bar v\leq0$ on $\mathbb R^n\setminus\mathcal X$.
Moreover, since $\mathcal X_b$ is closed and
$\mathcal X\subseteq\operatorname{int}(\mathcal X_b)$,
every boundary point of $\mathcal X_b$ belongs to
$\mathcal X_b\setminus\mathcal X$, where $v\leq0$.
Together with the lower semicontinuity of $v$ on
$\mathcal X_b$, this ensures that $\bar v$ is lower
semicontinuous on $\mathbb R^n$.

Finally, since $f(x,u)\in\mathcal X_b$ for all
$x\in\mathcal X$ and $u\in\mathcal U$, the extension
does not change any of the one-step expectations in
\eqref{eq:lsc-barrier-conditions}. It also preserves
boundedness on $\mathcal X$. Hence, the two certificate
domains yield equivalent characterizations.
\end{proof}

Theorem~\ref{thm:uniform-lsc} provides a necessary-and-sufficient
characterization of reach-avoid feasibility using a single
lower-semicontinuous certificate. An alternative
characterization, established in
\cite{xue2021reach,xue2026converse}, introduces an
additional auxiliary function and is given below.

\begin{proposition} \cite{xue2021reach,xue2026converse}
\label{prop:auxiliary-characterization}
    Under Assumptions \ref{assump1} and \ref{assum2},
$\mathcal X_0\subseteq\mathcal R$ holds if and only if 
there exist a bounded lower semicontinuous function $v:\mathcal{X}_b\rightarrow \mathbb{R}$ and a bounded lower semicontinuous function $w:\mathcal{X}_b\rightarrow \mathbb{R}$ satisfying 
\begin{equation}
\label{w}
    \begin{cases}
        v(x)\geq \epsilon,& \forall x\in \mathcal{X}_0,\\
        v(x)\leq  \mathbb{E}_{\nu}[v(f(x,\theta))],& \forall x\in \mathcal{X}\setminus \mathcal{T},\\
        v(x)\leq \mathbb{E}_{\nu}[w(f(x,\theta))]-w(x), & \forall x\in \mathcal{X}\setminus \mathcal{T},\\
        v(x)\leq 1, & \forall x\in \mathcal{T},\\
        v(x)\leq 0, & \forall x\in \mathcal{X}_b\setminus \mathcal{X}.
    \end{cases}
\end{equation}
\end{proposition}

The two characterizations offer different structures. Theorem~\ref{thm:uniform-lsc} uses a single
lower-semicontinuous function with a discounted
one-step inequality, whereas
Proposition~\ref{prop:auxiliary-characterization}
introduces an auxiliary function to characterize the
expected one-step increment of the certificate. For detailed comparisons of these two characterizations, please refer to \cite{cao2025comparative}.

\subsection{Continuous Lyapunov-Like Characterization}
\label{subsec:continuous-certificates}
The preceding subsection establishes a necessary-and-
sufficient reach-avoid characterization in terms of
lower-semicontinuous certificate functions. We now
investigate conditions under which such certificates
can be chosen continuous. This requires additional
regularity of the stochastic system induced by
randomizing the control inputs, which we formulate
in terms of its one-step transition kernel.

The randomized system induces the one-step transition
kernel
\begin{equation*}
    K_\nu(x,A)
    :=
    \nu\big(
        \{u\in\mathcal U \mid f(x,u)\in A\}
    \big),
    \qquad A\in\mathcal B(\mathcal X_b),
    \label{eq:transition-kernel}
\end{equation*}
for $x\in\mathcal X$, where
$\mathcal B(\mathcal X_b)$ denotes the Borel
$\sigma$-algebra on $\mathcal X_b$.

To obtain continuous certificates, we impose an
additional regularity condition on this family of
transition measures.

\begin{assumption}
\label{assump3}
The family of transition measures
$\{K_\nu(x,\cdot) \mid x\in\mathcal X\setminus\mathcal T\}$
is uniformly absolutely continuous with respect to the
$n$-dimensional Lebesgue measure $\mu$ on $\mathcal X_b$.
That is, for every $\rho>0$, there exists $\eta>0$
such that, for every Borel set $A\subseteq\mathcal X_b$,
\begin{equation}
    \mu(A)<\eta
    \quad\Longrightarrow\quad
    \sup_{x\in\mathcal X\setminus\mathcal T}
    K_\nu(x,A)<\rho.
    \label{eq:uniform-absolute-continuity}
\end{equation}
\end{assumption}
Assumption \ref{assump3} imposes a non-degeneracy requirement on the
effect of control inputs on the successor state. A useful sufficient condition can be
given for control-affine systems with equal control and
state dimensions. Consider
\[
    f(x,u)=f_0(x)+G(x)u,
\]
where \(G(x)\in\mathbb R^{n\times n}\). Suppose that
the control distribution \(\nu\) admits a bounded density
\(q\) with respect to Lebesgue measure on \(\mathcal U\),
and that the control-to-successor-state map is uniformly
nonsingular over \(x\in\mathcal X\setminus\mathcal T\),
namely,
\[
    \|q\|_{L^\infty(\mathcal U)}<\infty,
    \qquad
    \inf_{x\in\mathcal X\setminus\mathcal T}
    |\det G(x)|>0,
\]
where
\[
    \|q\|_{L^\infty(\mathcal U)}
    :=
    \operatorname*{ess\,sup}_{u\in\mathcal U}|q(u)|
\]
denotes the essential supremum of \(q\) with respect
to Lebesgue measure on \(\mathcal U\).
Under these conditions, the change-of-variables formula
implies that the transition kernel admits the density
\[
    p_\nu(y\mid x)
    =
    \frac{
        q\big(G(x)^{-1}(y-f_0(x))\big)
    }{|\det G(x)|},
\]
where the density is understood to be zero outside
\(f(x,\mathcal U)\). In particular, the transition
densities are uniformly bounded:
\[
    \sup_{x\in\mathcal X\setminus\mathcal T}
    \|p_\nu(\cdot\mid x)\|_{L^\infty(\mathcal X_b)}
    \leq
    \frac{\|q\|_{L^\infty(\mathcal U)}}
    {\displaystyle
    \inf_{x\in\mathcal X\setminus\mathcal T}
    |\det G(x)|}
    <\infty.
\]
Consequently, there exists a constant \(M>0\) such that
\[
    K_\nu(x,A)\leq M\mu(A),
    \qquad
    \forall x\in\mathcal X\setminus\mathcal T,\quad
    \forall A\in\mathcal B(\mathcal X_b).
\]
This uniform bound directly implies
Assumption~\ref{assump3}.

We are now ready to state the main result. Under
Assumption \ref{assump3}, the lower-semicontinuous characterization
can be strengthened to a continuous one, as established
by the converse result for stochastic discrete-time
systems in \cite{xue2026sufficient1}.

\begin{theorem}
\label{thm:main}
Suppose that Assumptions \ref{assump1}--\ref{assump3} hold. Then
$\mathcal X_0\subseteq\mathcal R$ if and only if there
exist a continuous function
$\widetilde v\in C(\mathcal X_b)$ and constants
$\varepsilon\in(0,1)$, $\gamma\in(0,1)$\footnote{Similarly, the parameter \(\gamma\) can be fixed a priori arbitrarily close to one without compromising the theoretical characterization.}, and
$\beta\geq0$ such that
\begin{equation}
\label{eq:continuous-barrier-conditions}
\begin{cases}
\displaystyle
    \widetilde v(x)
    \geq
    \varepsilon+
    \frac{\varepsilon\beta-\beta}{1-\gamma},
    & \forall x\in\mathcal X_0,\\[8pt]
\displaystyle
    \beta+\widetilde v(x)
    \leq
    \gamma\mathbb E_\nu[
        \widetilde v(f(x,\theta))
    ],
    & \forall x\in\mathcal X\setminus\mathcal T,\\[8pt]
    \widetilde v(x)\leq1,
    & \forall x\in\mathcal T,\\[4pt]
    (1-\gamma)\widetilde v(x)\leq-\beta,
    & \forall x\in\mathcal X_b\setminus\mathcal X.
\end{cases}
\end{equation}
Moreover, whenever $\mathcal X_0\subseteq\mathcal R$,
the parameters and the continuous function
$\widetilde v$ can be chosen such that all four
inequalities in \eqref{eq:continuous-barrier-conditions}
hold strictly.
\end{theorem}
\begin{proof}
Suppose first that there exist
$\widetilde v\in C(\mathcal X_b)$ and constants
$\varepsilon\in(0,1)$, $\gamma\in(0,1)$, and
$\beta\geq0$ satisfying
\eqref{eq:continuous-barrier-conditions}.
By the sufficiency of the continuous certificate
conditions established in \cite{xue2026sufficient1},
\[
    P_{\mathrm{RA}}(x)\geq\varepsilon,
    \qquad \forall x\in\mathcal X_0.
\]
Since $\varepsilon>0$, Proposition~\ref{prop:det-stoch-equivalence}
implies that $\mathcal X_0\subseteq\mathcal R$.

Conversely, suppose that $\mathcal X_0\subseteq\mathcal R$.
By Propositions~\ref{prop:det-stoch-equivalence}
and~\ref{prop:uniform-margin}, there exists
$\varepsilon\in(0,1)$ such that
\[
    P_{\mathrm{RA}}(x)>\varepsilon,
    \qquad \forall x\in\mathcal X_0.
\]
Under Assumptions \ref{assump1} and \ref{assump3}, the continuous converse
characterization in \cite{xue2026sufficient1}
guarantees the existence of
$\gamma\in(0,1)$, $\beta\geq0$, and
$\widetilde v\in C(\mathcal X_b)$ satisfying all four
inequalities in \eqref{eq:continuous-barrier-conditions}
strictly. This establishes the necessity.
\end{proof}

Theorem~\ref{thm:main} provides a necessary-and-sufficient
continuous Lyapunov-like characterization of controlled
discrete-time reach-avoid verification. It is obtained by
combining the probabilistic reformulation of controlled
reach-avoid feasibility with the continuous converse
characterization for stochastic systems. The continuous characterization also provides a foundation for computational verification. In particular, when the system dynamics and the sets involved admit suitable polynomial descriptions, \cite{xue2026sufficient1} also establishes the completeness of the corresponding sum-of-squares (SOS) approach under the stated algebraic assumptions. That is, whenever the reach-avoid specification is feasible, an SOS-based search is guaranteed to find a certificate of the prescribed form at a sufficiently high relaxation degree. This result connects the necessary-and-sufficient continuous characterization to a systematically convergent computational framework.
\section{Conclusion}
This paper introduced necessary-and-sufficient Lyapunov-like characterizations for reach-avoid
verification of controlled deterministic discrete-time systems. The approach relies on a probabilistic reformulation that
equates deterministic reach-avoid feasibility with
positive-probability reach-avoid under randomized controls.
Combining this equivalence with uniform probability margins
and converse characterizations for stochastic
systems yields the desired result.
\bibliographystyle{IEEEtran}
\bibliography{ref}

@article{wu2026controlled,
  title={Controlled Reach-avoid Set Computation for Discrete-time Polynomial Systems via Convex Optimization},
  author={Wu, Taoran and Xue, Yiling and Ren, Dejin and Easwaran, Arvind and Fr{\"a}nzle, Martin and Xue, Bai},
  journal={IEEE Transactions on Automatic Control},
  year={2026},
  publisher={IEEE}
}

@article{xue2026sufficient,
  title={Sufficient and necessary barrier-like conditions for safety and reach-avoid verification of stochastic discrete-time systems},
  author={Xue, Bai},
  journal={Automatica},
  volume={187},
  pages={112919},
  year={2026},
  publisher={Elsevier}
}

@article{xue2026converse,
  title={Converse barrier certificates for set-based stochastic reach-avoid verification},
  author={Xue, Bai and Ong, C-H Luke},
  journal={arXiv preprint arXiv:2608.30318},
  year={2026}
}

@article{xue2026sufficient1,
  title={Sufficient and necessary continuous barrier-like conditions for discrete-time stochastic reach-avoid verification},
  author={Xue, Bai},
  journal={arXiv preprint arXiv:2609.23973},
  year={2026}
}

@inproceedings{xue2021reach,
  title={Reach-avoid analysis for stochastic discrete-time systems},
  author={Xue, Bai and Li, Renjue and Zhan, Naijun and Fr{\"a}nzle, Martin},
  booktitle={2021 American Control Conference (ACC)},
  pages={4879--4885},
  year={2021},
  organization={IEEE}
}

@book{khalil2002nonlinear,
  title={Nonlinear systems},
  author={Khalil, Hassan K and Grizzle, Jessy W},
  volume={3},
  year={2002},
  publisher={Prentice hall, Upper Saddle River, NJ}
}

@article{prajna2007framework,
  title={A framework for worst-case and stochastic safety verification using barrier certificates},
  author={Prajna, Stephen and Jadbabaie, Ali and Pappas, George J},
  journal={IEEE Transactions on Automatic Control},
  volume={52},
  number={8},
  pages={1415--1428},
  year={2007},
  publisher={IEEE}
}

@article{prajna2007convex,
  title={Convex programs for temporal verification of nonlinear dynamical systems},
  author={Prajna, Stephen and Rantzer, Anders},
  journal={SIAM Journal on Control and Optimization},
  volume={46},
  number={3},
  pages={999--1021},
  year={2007},
  publisher={SIAM}
}

@inproceedings{papachristodoulou2002construction,
  title={On the construction of Lyapunov functions using the sum of squares decomposition},
  author={Papachristodoulou, Antonis and Prajna, Stephen},
  booktitle={Proceedings of the 41st IEEE Conference on Decision and Control, 2002.},
  volume={3},
  pages={3482--3487},
  year={2002},
  organization={IEEE}
}

@inproceedings{Tomlin1999Computing,
  author    = {Claire J. Tomlin and John Lygeros and S. Shankar Sastry},
  title     = {Computing Controllers for Nonlinear Hybrid Systems},
  booktitle = {Hybrid Systems: Computation and Control},
  series    = {Lecture Notes in Computer Science},
  volume    = {1569},
  pages     = {238--255},
  publisher = {Springer, Berlin, Heidelberg},
  year      = {1999},
  doi       = {10.1007/3-540-48983-5_22}
}

@article{margellos2011hamilton,
  title={Hamilton--Jacobi formulation for reach--avoid differential games},
  author={Margellos, Kostas and Lygeros, John},
  journal={IEEE Transactions on automatic control},
  volume={56},
  number={8},
  pages={1849--1861},
  year={2011},
  publisher={IEEE}
}

@article{liu2021converse,
  title={Converse barrier functions via Lyapunov functions},
  author={Liu, Jun},
  journal={IEEE Transactions on Automatic Control},
  volume={67},
  number={1},
  pages={497--503},
  year={2021},
  publisher={IEEE}
}

@inproceedings{quartz2026converse,
  title={A converse control Lyapunov theorem for joint safety and stability},
  author={Quartz, Thanin and Fitzsimmons, Maxwell and Liu, Jun},
  booktitle={2026 American control conference (ACC)},
  pages={3739--3744},
  year={2026},
  organization={IEEE}
}

@article{mestres2025converse,
  title={Converse theorems for certificates of safety and stability},
  author={Mestres, Pol and Cort{\'e}s, Jorge},
  journal={IEEE Transactions on Automatic Control},
  volume={70},
  number={12},
  pages={8016--8028},
  year={2025},
  publisher={IEEE}
}

@article{li2026converse,
  title={Converse barrier certificates for finite-time safety verification of continuous-time perturbed deterministic systems},
  author={Li, Yonghan and Wu, Chenyu and Wu, Taoran and Wang, Shijie and Xue, Bai},
  journal={Systems \& Control Letters},
  volume={209},
  pages={106357},
  year={2026},
  publisher={Elsevier}
}

@article{lin1996smooth,
  title={A smooth converse Lyapunov theorem for robust stability},
  author={Lin, Yuandan and Sontag, Eduardo D and Wang, Yuan},
  journal={SIAM Journal on Control and Optimization},
  volume={34},
  number={1},
  pages={124--160},
  year={1996},
  publisher={SIAM}
}

@article{ratschan2018converse,
  title={Converse theorems for safety and barrier certificates},
  author={Ratschan, Stefan},
  journal={IEEE Transactions on Automatic Control},
  volume={63},
  number={8},
  pages={2628--2632},
  year={2018},
  publisher={IEEE}
}

@article{majumdar2025sound,
  title={Sound and complete proof rules for probabilistic termination},
  author={Majumdar, Rupak and Sathiyanarayana, VR},
  journal={Proceedings of the ACM on Programming Languages},
  volume={9},
  number={POPL},
  pages={1871--1902},
  year={2025},
  publisher={ACM, New York, NY, USA}
}

@article{cao2025comparative,
  title={Comparative analysis of barrier-like function methods for reach-avoid verification in stochastic discrete-time systems},
  author={Cao, Zhipeng and Wang, Peixin and Ong, Luke and {\v{Z}}ikeli{\'c}, {\DJ}or{\dj}e and Wagner, Dominik and Xue, Bai},
  journal={arXiv preprint arXiv:2512.05348},
  year={2025}
}
\end{document}